\documentclass[prl,aps,amssymb,showpacs,footinbib,preprint,superscriptaddress,times]{revtex4-1}

\usepackage[english]{babel}
\usepackage[pdftex,colorlinks=true,linkcolor=blue,citecolor=blue,urlcolor=blue]{hyperref} 
\usepackage{amsmath,amssymb,amsthm}

\usepackage{amsmath,amssymb,mathrsfs}
\usepackage{latexsym}
\usepackage{graphicx}
\usepackage{epstopdf}
\usepackage{color}
\usepackage{amsfonts}
\usepackage{bm}
\usepackage{bbm}
\usepackage{multirow}
\usepackage{subfiles}

\usepackage{natbib}

\usepackage{lipsum}
\usepackage{lineno}

\usepackage{ulem} 

\usepackage{tikz} 
\usepackage{tikz-feynman}

\usepackage{mathtools} 
\usepackage{physics}
\usepackage[version=3]{mhchem}

\usepackage{microtype} 
\usepackage{csquotes} 
\usepackage{booktabs} 

\makeatletter
\renewcommand\@make@capt@title[2]{%
	\@ifx@empty\float@link{\@firstofone}{\expandafter\href\expandafter{\float@link}}%
	{\textbf{#1}}\@caption@fignum@sep#2\quad
}%
\makeatother

\makeatletter
\g@addto@macro\bfseries{\boldmath} 
\makeatother

\usepackage{graphicx}
\usepackage{placeins}

\usepackage{siunitx} 
\DeclareSIUnit\year{a}
\DeclareSIUnit\pixel{px}
\DeclareSIUnit\line{line}

\usepackage{scrlayer-scrpage} 
\usepackage{lastpage} 
\usepackage{placeins}
\usepackage{layouts}

\newcommand{\cre}{{\dag}}
\newcommand{\ann}{{\vphantom{\dag}}}

\graphicspath{./figures/}

\usepackage{float}
\def\bra#1{\left\langle#1\right|}
\def\ket#1{\left|#1\right\rangle}

\newcommand\boldvector[1]{%
  \ifcat\noexpand#1\relax 
    \boldsymbol{#1}
  \else
    \mathbf{#1}
  \fi
}

\def\frac#1#2{{\textstyle{#1 \over #2}}}

\newcommand{\rom}[1]{\uppercase\expandafter{\romannumeral #1\relax}}

\begin{document}


\title{Pomeranchuk instability from electronic correlations in CsTi$_3$Bi$_5$ kagome metal}



\author{Chiara Bigi**}
\affiliation{Synchrotron SOLEIL, L’Orme des Merisiers, D\'epartementale 128, F-91190 Saint-Aubin, France}

\author{Matteo D\"urrnagel**}
\affiliation{Institute for Theoretical Physics and Astrophysics, University of W\"urzburg, D-97074 W\"urzburg, Germany}
\affiliation{Institute for Theoretical Physics, ETH Z\"urich, 8093 Z\"urich, Switzerland}

\author{Lennart Klebl}
\affiliation{Institute for Theoretical Physics and Astrophysics, University of W\"urzburg, D-97074 W\"urzburg, Germany}
\affiliation{I. Institute for Theoretical Physics, Universit\"at Hamburg, Notkestrasse 9-11, 22607 Hamburg, Germany}

\author{Armando Consiglio}
\affiliation{Istituto Officina dei Materiali, Consiglio Nazionale delle Ricerche, Trieste I-34149, Italy}

\author{Ganesh Pokharel}
\affiliation{Materials Department, University of California Santa Barbara, Santa Barbara, California 93106, USA}
\affiliation{Perry College of Mathematics, Computing, and Sciences, University of West Georgia, Carrollton, GA 30118, USA}

\author{Fran\c cois Bertran}
\affiliation{Synchrotron SOLEIL, L’Orme des Merisiers, D\'epartementale 128, F-91190 Saint-Aubin, France}

\author{Patrick Le F\`evre}
\affiliation{Univ Rennes, IPR Institut de Physique de Rennes, UMR 6251, F-35000 Rennes, France}

\author{Thomas Jaouen}
\affiliation{Univ Rennes, IPR Institut de Physique de Rennes, UMR 6251, F-35000 Rennes, France}

\author{Hulerich C. Tchouekem}
\affiliation{Univ Rennes, IPR Institut de Physique de Rennes, UMR 6251, F-35000 Rennes, France}

\author{Pascal Turban}
\affiliation{Univ Rennes, IPR Institut de Physique de Rennes, UMR 6251, F-35000 Rennes, France}

\author{Alessandro De Vita}
\affiliation{Fritz Haber Institut der Max Planck Gesellshaft, Faradayweg 4--6, 14195 Berlin, Germany\looseness=-1}

\author{Jill A. Miwa}
\affiliation{Department of Physics and Astronomy, Interdisciplinary Nanoscience Center, Aarhus University, 8000 Aarhus C, Denmark}

\author{Justin W. Wells}
\affiliation{Department of Physics and Centre for Materials Science and Nanotechnology, University of Oslo (UiO), 0318 Oslo, Norway.}

\author{Dongjin Oh}
\affiliation{Department of Physics, Massachusetts Institute of Technology, Cambridge, MA, USA}

\author{Riccardo Comin}
\affiliation{Department of Physics, Massachusetts Institute of Technology, Cambridge, MA, USA}

\author{Ronny Thomale}
\affiliation{Institute for Theoretical Physics and Astrophysics, University of W\"urzburg, D-97074 W\"urzburg, Germany}

\author{Ilija Zeljkovic}
\affiliation{Department of Physics, Boston College, Chestnut Hill, MA 02467, USA}

\author{Brenden R. Ortiz}
\affiliation{Materials Department, University of California Santa Barbara, Santa Barbara, California 93106, USA}

\author{Stephen D. Wilson}
\affiliation{Materials Department, University of California Santa Barbara, Santa Barbara, California 93106, USA}

\author{Giorgio Sangiovanni}
\affiliation{Institute for Theoretical Physics and Astrophysics, University of W\"urzburg, D-97074 W\"urzburg, Germany}

\author{Federico Mazzola}\email{federico.mazzola@spin.cnr.it}
\affiliation{CNR-SPIN, c/o Complesso di Monte S. Angelo, IT-80126 Napoli, Italy}
\affiliation{Department of Molecular Sciences and Nanosystems,
Ca’ Foscari University of Venice, 30172 Venice, Italy}

\author{Domenico Di Sante}\email{domenico.disante@unibo.it}
\affiliation{Department of Physics and Astronomy, University of Bologna, 40127 Bologna, Italy}

\date{\today}


\begin{abstract}
\textbf{Among many-body instabilities in correlated quantum systems, electronic nematicity, defined by the spontaneous breaking of rotational symmetry, has emerged as a critical phenomenon, particularly within high-temperature superconductors. Recently, this behavior has been identified in CsTi$_3$Bi$_5$, a member of the AV$_3$Sb$_5$ (A = K, Rb, Cs) kagome family, recognized for its intricate and unconventional quantum phases. Despite accumulating indirect evidence, the fundamental mechanisms driving nematicity in CsTi$_3$Bi$_5$ remain inadequately understood, sparking ongoing debates. In this study, we employ polarization-dependent angle-resolved photoemission spectroscopy to reveal definitive signatures of an orbital-selective nematic deformation in the electronic structure of CsTi$_3$Bi$_5$. This direct experimental evidence underscores the pivotal role of orbital degrees of freedom in symmetry breaking, providing new insights into the complex electronic environment. By applying the functional renormalization group technique to a fully interacting ab initio model, we demonstrate the emergence of a finite angular momentum ($d$-wave) Pomeranchuk instability in CsTi$_3$Bi$_5$, driven by the concomitant action of electronic correlations within specific orbital channels and chemical potential detuning away from Van Hove singularities. By elucidating the connection between orbital correlations and symmetry-breaking instabilities, this work lays a crucial foundation for future investigations into the broader role of orbital selectivity in quantum materials, with far-reaching implications for the design and manipulation of novel electronic phases.}
\end{abstract}

\maketitle

** These authors contributed equally\\

Kagome metals represent a remarkable class of materials that has garnered significant attention in condensed matter physics due to the richness of observed correlated phases.
Owing to the unique topological properties of the underlying kagome lattice, highly itinerant electrons naturally display a variety of massless Dirac-like states, Van Hove singularities (VHss) with partial sublattice polarisation~\cite{Kiesel2012, Hu2022, Kang2022}, and compact localised states with dispersionless bands within a single band structure~\cite{Kang2020, Li2021, Liu2020, DiSante2023}.
Combined with the inherent geometrical frustration of any interacting model on the kagome lattice, this provides ideal conditions for the realisation of many exotic phases long sought in other correlated electron platforms including persistent loop current formation
~\cite{Mielke2022, Yu2021, Guo2022, Neupert2022} and superconducting pairing modulations~\cite{Chen2021, Zhou2022, Schwemmer2024, Deng2024}.
While the experimental evidence for these phases is still subject of ongoing debates, electronic nematicity, i.e. the breaking of rotational symmetry in the charge ordered state, strikes out as one of the few universal features across all kagome compounds~\cite{Jiang2021, Yin2022, Xu2022, Tuniz2023, Hu2023, Jiang2024, Drucker2024, Nag2024}.
Despite its critical relevance also for the subsequent superconducting transition at low temperature~\cite{Boehmer2022, kang2023, Le2024, Deng2024}, the origin and microscopic mechanism governing nematicity has remained elusive, necessitating focused investigation.

Among the diverse array of kagome systems, the first synthesized AV$_3$Sb$_5$ family (where A = K, Rb, Cs)~\cite{Ortiz2019, Ortiz2020} has remained the most extensively studied.
However, the simultaneous emergence and intertwining of the various correlated electronic phases induced by a lack of scale separation between different symmetry breaking phases poses a significant challenge to the understanding of the underlying mechanisms driving the various many-body effects and remains a current thread to any thorough assessment of kagome metals~\cite{Zhao_2021, Zheng2022}.

In this study, we utilize \ce{CsTi3Bi5} as a unique platform to explore the phenomenon of rotational symmetry breaking on the kagome lattice.
The absence of an accompanying translational symmetry breaking typical of other '135' kagome metals renders this compound an ideal experimental setting for a detailed study of nematicity~\cite{Li2023, Jiang2023, Hu2023}.
To elucidate the underlying mechanisms, we adopt a synergistic approach that combines light-polarization-dependent angle-resolved photoelectron spectroscopy (ARPES) with ab initio based field theoretical methods.
Our findings reveal that electronic nematicity in \ce{CsTi3Bi5} reduces the symmetry of the system through an orbital-selective mechanism, with dominant contributions from the planar $d_{xy}$ and $d_{x^2-y^2}$ orbitals of \ce{Ti}.
Crucially, our results indicate a purely electronic origin of this effect, stemming from the frustrated long-range Coulomb repulsion and considerable VHs detuning from the Fermi level.
This supports the existence of a $d$-wave Pomeranchuk instability (PI) in \ce{CsTi3Bi5}, characterized by the spontaneous breaking of point group symmetry driven by the divergence of an associated susceptibility of the electronic system~\cite{Pomeranchuk1958}, and a sublattice charge imbalance. In this respect, our findings notably differ from the prevailing view that nematic charge order in titanium-based kagome metals arises through a bond-type order with zero total momentum~\cite{Hu2023,Li2023}. That interpretation originated from observations of the momentum-dependent character of nematic Fermi surface distortions, which were incorrectly claimed to be incompatible with a straightforward charge imbalance across unit cell sites~\cite{Hu2023}, as well as from many-body calculations that favored non-local charge orders~\cite{Huang2024}.


%


Recently, PIs have sparked recurrent interest. In the context of multi-layer graphene~\cite{Saito2021, Rozen2021}, the PI is realized as a partial valley polarisation and results in exotic half and quarter-metal states where the iso-spin provides the additional ingredient to realize the spontaneous rotational symmetry breaking~\cite{Chichinadze2020}.
The surface states of topological elemental arsenic ($\alpha$-As) have now also been reported to support a genuine PI~\cite{Hossain2024}, despite weak correlations. Our work on CsTi$_3$Bi$_5$, on the other hand, puts forward PI as a generic instability in hexagonal lattice systems with a non-trivial sublattice degree of freedom, advancing our understanding about the microscopic mechanisms governing electronic nematicity.

\begin{figure*}[t]
\centering
\includegraphics[width=1.\textwidth]{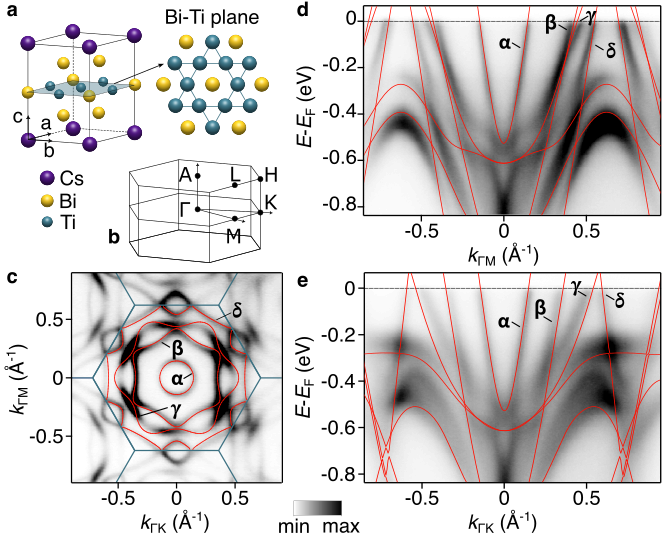}
\caption{\textbf{Crystal and electronic structure of CsTi$_3$Bi$_5$.} \textbf{a}, Cartoon of the unit cell structure. The shaded area marks the Bi-Ti plane where the Ti atoms form a kagome lattice. \textbf{b}, Brillouin zone with high symmetry points. \textbf{c}, Fermi surface collected at the Brillouin zone centre with h$\nu$ = 65 eV circularly polarised light (summing up right- and left-handed polarisations). \textbf{d}, Energy versus momentum dispersion along $\Gamma$-M and \textbf{e}, $\Gamma$-K high-symmetry directions. Bulk, non-nematic first-principles band structures (red solid lines) are superimposed to the experimental data in panels \textbf{c}, \textbf{d} and \textbf{e}.}
\label{fig1}
\end{figure*}

CsTi$_3$Bi$_5$ exemplifies the defining structure of the '135' kagome metals, being isostructural with the extensively studied V-based compounds AV$_3$Sb$_5$. Within the Bi-Ti plane, the kagome lattice formed by Ti atoms (Fig.\ref{fig1}a) is pivotal in establishing a complex electronic structure characterized by itinerant Dirac-like states, VHss, and flat bands~\cite{Yi2023, Yang2023, Wang2023, Huang2024}. Our angle-resolved photoemission spectroscopy (ARPES) measurements unveil sharply defined bands that disperse across the Fermi level, underpinning the metallic nature of the band structure and resulting in a multifaceted Fermi surface (Fig.\ref{fig1}c) defined by multiple sheets with an overall hexagonal geometry. Importantly, the experimental spectra illustrated in Fig.\ref{fig1}d,e exhibit concordance with predictions from first-principles calculations, that is even enhanced when taking into account surface termination effects (Supplementary Information I), thereby reinforcing the reliability of our findings. We emphasize that these data are obtained through the integration of ARPES spectra collected using both circular-right and circular-left polarizations, ensuring comprehensive resolution of the principal electronic features. Additional spectra recorded with linear vertical and horizontal polarizations, along with varying photon energies, are provided in \ref{ExtDataFig1} and \ref{ExtDataFig1bis}. Notably, the electronic band structure of CsTi$_3$Bi$_5$ reveals a pronounced absence of $k_z$ dispersion for the dispersion at the energy of the Fermi level, with no significant variations detected upon photon energy adjustment (also consistent with the almost cylindrical ab initio Fermi surface shown in Fig.\ref{fig3}c), thereby underscoring its distinct two-dimensional character.

\begin{figure*}[t]
\centering
\includegraphics[width=1.\textwidth]{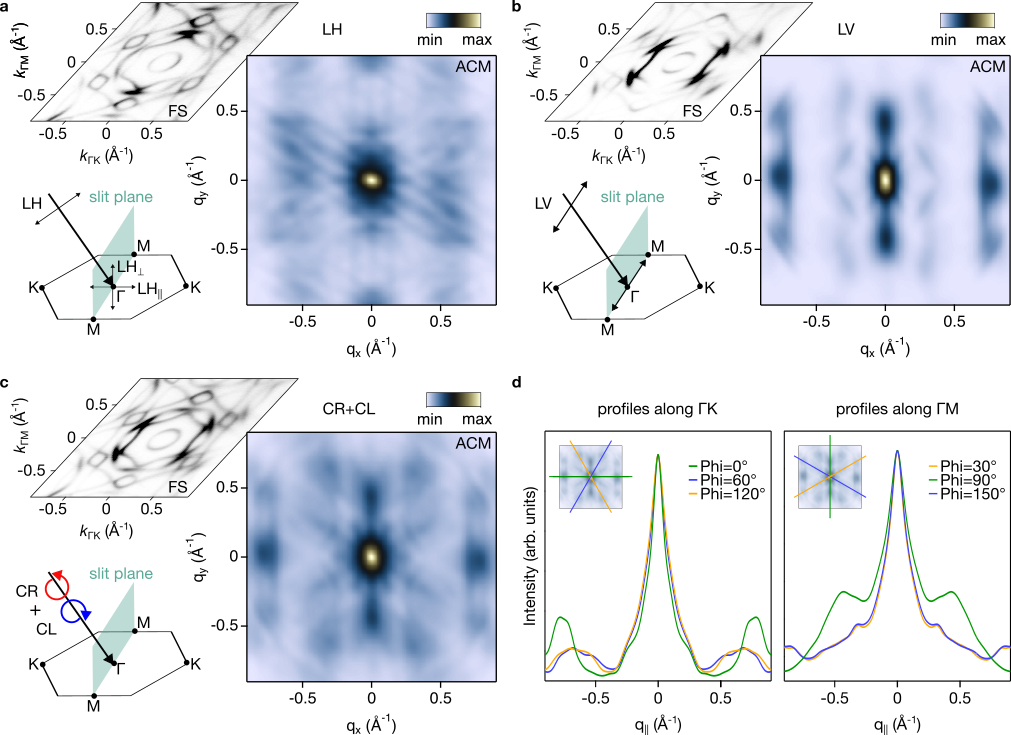}
\caption{\textbf{Electronic nematicity of CsTi$_3$Bi$_5$ Fermi contour.} \textbf{a}, Autocorrelation map (ACM) built from the Fermi surface contour measured at the Brillouin zone centre with h$\nu$ = 65 eV for linear horizontal polarised light and the analyser slit aligned along the $\Gamma$-M high-symmetry direction. The cartoon sketches the experimental geometry and vector projections. \textbf{b}, Same as \textbf{a}, but for linear vertical polarised light, to rule out geometrical and matrix element effects. \textbf{c}, Same as \textbf{a} and \textbf{b}, but for unpolarised light (summing up spectra collected with right- and left-handed circularly polarised light). \textbf{d}, Azimuthal profiles extracted from the ACM in \textbf{c}, supporting the reduced nematic C$_2$ symmetry.}
\label{fig2}
\end{figure*}

\begin{figure*}[t]
\centering
\includegraphics[width=1.\textwidth]{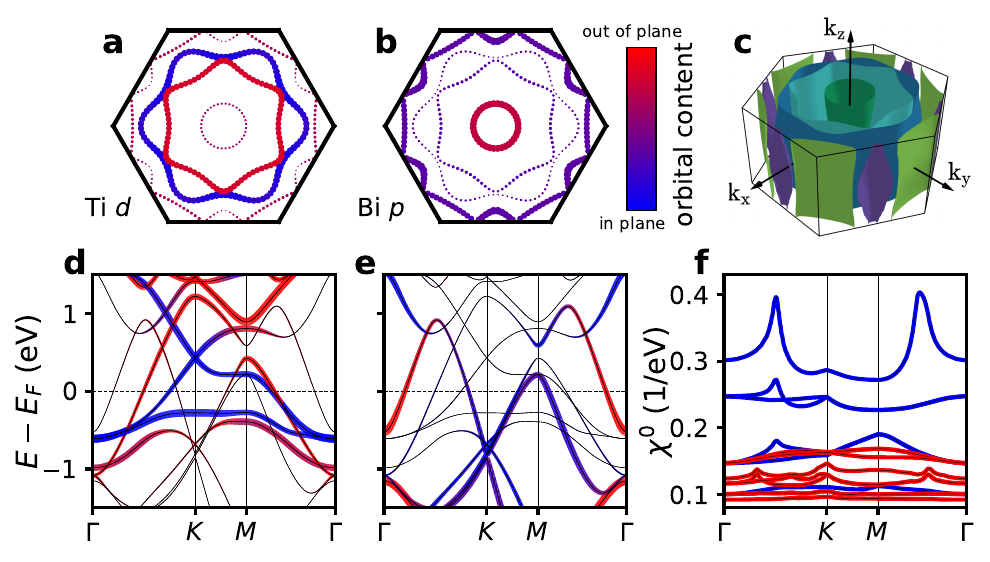}
\caption{\textbf{Orbital polarisation of the Fermi sheets.} \textbf{a},\textbf{b}, Integrated orbital content on the $k_z = 0$ Fermi surface for the Ti $d$-orbitals and Bi $p$-orbitals, as indicated by dot size and color. The two inner Fermi pockets $(\beta)$ and $(\gamma)$ show clear orbital polarisation with respect to in-plane and out-of-plane alignment of the $d$-orbitals.
\textbf{c}, The three-dimensional Fermi surface shows little $k_z$ dispersion.
\textbf{d},\textbf{e}, The orbital character of the full non-interacting dispersion for Ti $d$ and Bi $p$, revealing a low-lying Van Hove singularity that is exclusively supported by the in-plane $d$-orbitals. \textbf{f}, This is directly reflected in the pronounced role of in-plane orbital combinations in the components of the orbital-resolved bare electronic susceptibility $\chi^0$ given by Eq.~\eqref{eqn:bare_susc}.}
\label{fig3}
\end{figure*}

Significant spectroscopic variations emerge when light polarization is altered, particularly between linear horizontal and vertical orientations. The horizontal configuration, with a 45-degree incidence angle, incorporates both in-plane and out-of-plane components. In contrast, the vertical orientation lies entirely within the crystal plane, aligned parallel to the analyzer slit along the $\Gamma$-M line. Schematic representations of this geometric arrangement are illustrated in Fig.\ref{fig2}a,b. The intensity of the photoemission process is directly related to the probability of a transition from an initial state to a final state. This probability, represented by the square module of the dipole matrix elements as $|\bra{\Psi^{k}_{fin}}\Vec{A} \cdot \Vec{p}\ket{\Psi^{k}_{in}}|^2$, is non-zero only when the symmetry of $\Vec{A} \cdot \Vec{p}\ket{\Psi^{k}_{in}}$ coincides with that of the final state. The detector, corresponding to a plane wave of even parity ($\ket{+}$), permits only even-parity transitions from the initial state, specifically $+\ket{+}$ and $-\ket{-}$. Conversely, transitions involving odd-parity initial state symmetry are forbidden ($\pm\ket{\mp}$), resulting in a null photoemission intensity signal~\cite{Damascelli2004, Damascelli2003}.

Understanding these polarization-dependent effects allows us to elucidate the orbital contributions to the electronic structure of CsTi$_3$Bi$_5$. This distinction is evident in the varying photoemission intensities observed for linear horizontal and vertical polarizations, as depicted in Fig.\ref{fig2}a and Fig.\ref{fig2}b, that highlight contrasting spectroscopic features. To enhance visualization, we also analyze the difference between the two polarizations (see \ref{ExtDataFig2}), providing insights into the relative contributions of in-plane and perpendicular orbitals. Our findings reveal that the inner circular pocket $(\alpha)$ centered at $\Gamma$ is predominantly composed of orbitals with out of plane symmetry. Additionally, the outermost hexagonal Fermi surface sheet $(\delta)$, forming the pockets around both the M and K points, exhibits similar characteristics.

In contrast, the two internal hexagonal Fermi surface sheets $(\beta)$ and $(\gamma)$, rotated by 30 degrees relative to one another, exhibit a substantial contribution from in-plane orbitals. To further elucidate the relative contributions of in-plane and out-of-plane orbitals while mitigating potential matrix element effects, we also performed light-polarization-dependent ARPES measurements across various experimental geometries, including rotations of the analyzer slit along the $\Gamma$-K direction. The response of the two hexagonal Fermi surface sheets to this rotation is notably distinct. The external $(\gamma)$ sheet retains a consistent in-plane orbital contribution, with its intensity adapting in accordance with the sample orientation. In contrast, the internal $(\beta)$ sheet demonstrates a shift in its spectral weight contribution, indicating a more complex orbital character. This behavior suggests that the external hexagonal $(\gamma)$ sheet is primarily composed of in-plane orbitals, while the internal $(\beta)$ sheet possesses a mixed orbital character. Support for this interpretation arises from the orbitally projected ab initio calculations shown in Fig. \ref{fig3}a, which assign $(d_{xy}, d_{x^2-y^2})$ orbitals to the external $(\gamma)$ sheet and a predominance of $(d_{xz}, d_{yz})$ orbitals to the internal one $(\beta)$. Remarkably, this conclusion is consistent across multiple experimental configurations. Additionally, our use of varied light polarizations, different photon energies, and the absence of significant $k_z$ dispersion, further confirm that our findings are robust against photoemission matrix element effects.

We now focus on the nematic character of the system, which has been previously reported by STM experiments~\cite{Li2023} to lead to a reduction of symmetry from $C_6$ to $C_2$. This alteration signifies that one of the three equivalent high-symmetry directions along both the $\Gamma$-K and $\Gamma$-M axes becomes distinct in terms of momentum length.
To investigate this phenomenon, we perform Lorentzian fits to the momentum distribution curves at the Fermi level for inequivalent $\Gamma$-M directions. Our analysis reveals a subtle variation in the momentum positions of the bands that constitute the Fermi surface, primarily influenced by in-plane components (see \ref{ExtDataFig3}). However, the observed difference is minimal, and given the momentum resolution, it remains challenging to definitively attribute these variations to nematicity. Such effects are anticipated to be subtle, as highlighted in previous works~\cite{Hu2023, Li2023}.

Autocorrelation maps (ACMs) of the Fermi surface (see Fig.\ref{fig2}) instead yield valuable insights into the symmetries inherent in the electronic structure, often elucidating even the most subtle features. \ce{CsTi3Bi5} serves as an ideal platform for this type of investigation, owing to the large extension of its nematic domains~\cite{Li2023}. By generating ACM under varying light polarizations, we are able to uncover the nematic $C_2$ symmetry of the sample, independent of light polarization. This is clearly demonstrated in Fig.\ref{fig2}d, where profiles along the three $\Gamma$-M and $\Gamma$-K directions are presented. These profiles distinctly reveal that two directions remain equivalent, while the third exhibits a marked difference. The $q$-vector extracted from both the $\Gamma$-M and $\Gamma$-K directions in the ACM corresponds to scattering within the $(\gamma)$ Fermi surface sheet, which is primarily composed of in-plane orbitals.
Further corroboration of our findings comes from conducting the same analysis after rotating the sample (see \ref{ExtDataFig4}), or changing the photon energy (see \ref{ExtDataFig6}). Despite certain shape differences likely attributable to matrix element effects, two profiles consistently maintain equivalence while one consistently differs, thereby reinforcing the reduced $C_2$ symmetry of the system.

The subtle signatures of rotational symmetry breaking presented above hinder a definite determination of the nematic order parameter based solely on the experimental data. The latter, in fact, points towards the direction that the system is nematic: based on autocorrelation maps showing a $C_2$ symmetry and on electronic structure momentum distribution curves fitting (See extended data), strong hints about this putative reduction are demonstrated. However, for substantial proof, it necessitates a detailed theoretical analysis on a realistic model to distinguish among the possible underlying microscopic origins~\cite{Fernandes2014}.
In other '135' kagome compounds like \ce{AV3Sb5}, nematicity is believed to emerge from a coupling between the different translation symmetry breaking orders at the inequivalent M points, that favors an imbalance between them to minimize the free energy within the symmetry broken phase~\cite{Nie2022, Neupert2022} and is highly prone to external effects like in-plane strain and magnetic field~\cite{Guo2024}.
The absence of a $2 \times 2$ superlattice modulation in the charge ordered phase of \ce{CsTi3Bi5} asks for a different explanation.
Likewise, \ce{CsTi3Bi5} does not feature unstable phonon modes and the electron-phonon coupling is negligible~\cite{Yi2023}. Since phonons are not expected to drive a nematic instability in $d=3$ dimensions~\cite{Cichutek2022} and hence seem not able to explain the accumulated experimental evidence, an exclusive electronic mechanism seems to drive the nematic transition in \ce{CsTi3Bi5}. 

To achieve a detailed understanding of the nematic transition and pin down its order parameter, we supplement the full ab initio dispersion (see Methods) with a electron-electron interactions between the \ce{Ti} $d$-orbitals via a site-local Kanamori vertex and density-density interactions up to second-nearest-neighbor distance.
We employ the FRG~\cite{Salmhofer2001, Metzner2012, Platt2013, Dupuis2021} in the timely truncated unity formulation (see Refs.~\cite{Beyer2022, Husemann2009, Lichtenstein2017, Profe2022, Profe2024a} and Methods) to obtain possible many-body instabilities of the system in an unbiased manner.

The large number of orbitals required to capture the low energy manifold is a common thread to any theoretical assessment of kagome metals. Hence, extracting the relevant degrees of freedom from the ab initio band structure is highly desirable to capture the kinetic theory in sophisticated many-body techniques.
In the orbital-resolved band structure of Fig.\ref{fig3}d,e the three characteristic kagome bands featuring a flat band, a Dirac cone at K and VHss at M are clearly visible. This low energy kagome manifold is exclusively supported by the in-plane $d$-orbitals of \ce{Ti}, while the out-of-plane \ce{Ti} $d$ and \ce{Bi} $p$ orbitals only contribute to bands with small spectral weight at the Fermi level.
This orbital hierarchy is directly reflected in the static response of electronic states to quantum fluctuations of a certain orbital structure, that can be quantified by the orbital resolved bare susceptibility tensor
\begin{equation} \label{eqn:bare_susc}
    \chi^0_{o_1o_2o_3o_4}(\mathbf{q}) =
    - \frac{1}{\beta} \sum_n
    \int_\text{BZ}\,\frac{\text{d}\mathbf k}{V_\text{BZ}}
    G_{o_2o_4}(\mathbf{k}, \omega_n)
    G_{o_3o_1}(\mathbf{k} + \mathbf{q}, \omega_n)
    + \text{h.c.} \, ,
\end{equation}
where $G_{o_1o_2}(\mathbf k, \omega_n)$ is the single-particle propagator with momentum $\mathbf k$, fermionic Matsubara frequency $\omega_n$ and orbital quantum numbers $o_i$, and the integral over momentum and frequency is normalized by the Brillouin zone volume $V_\text{BZ}$ and inverse temperature $\beta$, respectively.
The physical susceptibilities, \textit{i.e.} $\chi^0$ with $o_1 = o_2$, $o_3 = o_4$, are depicted in Fig.\ref{fig3}f.
The effect of the isolated VHs  of pure ($p$-type) sublattice character~\cite{Kiesel2012,Kiesel2013} in the vicinity of the Fermi level (Fig.~\ref{fig3}d) is directly apparent: while the peak in the susceptibility is shifted from the Van Hove scattering vector M closer to $\Gamma$ due to the detuning of the chemical potential from the perfect Van Hove filling, the in-plane contributions to $\chi^0$ (blue) dominate over out-of-plane parts (red).
Hence in-plane (anti-)screening processes play the decisive role in the determination of the system's ordering propensities and the emergence of a symmetry broken phase.

\begin{figure*}[t]
\centering
\includegraphics[width=1.\textwidth]{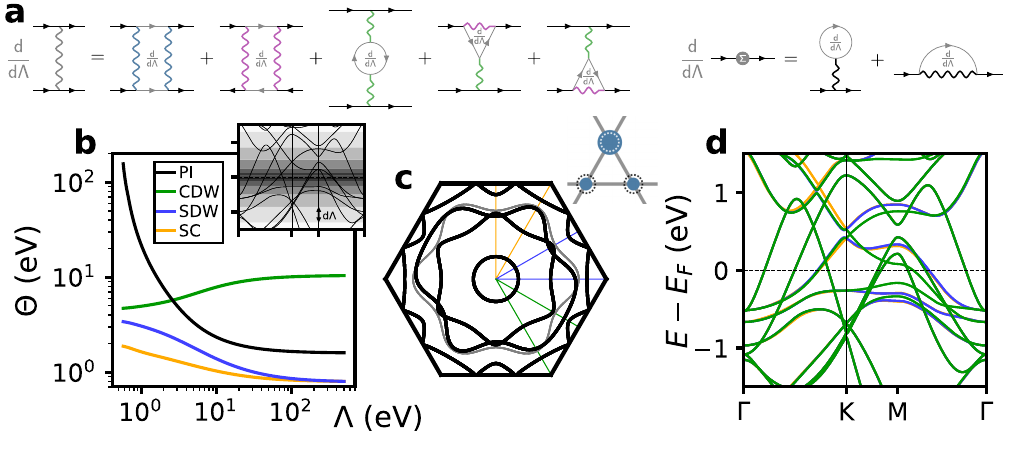}
\caption{\textbf{Orbital selective nematicity from electronic correlations}
\textbf{a}, FRG flow equations for the two-particle interaction vertex $V$ (left) and the self-energy $\Sigma$ (right). The different colored interaction lines indicate the transfer momenta characterising the diagrammatic channels as particle-particle (P, blue), direct (D, green) and crossed (C, purple) particle-hole. Gray objects represent scale derivatives $\mathrm{d}/\mathrm{d}\Lambda$.
\textbf{b}, Eigenvalues of the scattering vertex $\Theta$ for superconductivity (SC), $2 \times 2$ charge density wave (CDW), magnetic order spin density wave (SDW) and Pomeranchuk instability (PI).
The inset visualizes the successive integration of electronic modes in the energy range $\abs{E} \in [\Lambda, \Lambda + \text{d} \Lambda]$ during each step of the FRG flow.
\textbf{c}, Nematic Fermi surface reconstruction. The $C_6$ symmetry of the original Fermi surface (grey) is lowered to $C_2$ in the symmetry broken phase (black) without any backfolding.
The inset is a sketch of the corresponding charge distribution from the in-plane orbitals on the different kagome sublattice sites.
\textbf{d}, Band structure along the three formerly equivalent high-symmetry paths indicated in \textbf{c} in the nematic phase. Due to the broken $C_3$ symmetry by which $C_6$ is lowered to $C_2$, the Dirac cone is no longer pinned to the $K$ point.}
\label{fig4}
\end{figure*}

We therefore evaluate the FRG flow equations for the two-particle vertex (lhs.~of Fig.\ref{fig4}a) with the in-plane orbitals as interacting subspace.
We uncover a $Q = 0$ divergence in the charge susceptibility, that corresponds to an intra-unit cell charge density wave order transforming under the two dimensional $E_2$ irreducible representation of the crystalline point group $P6/mmm$.
In addition to the two-particle interaction, we also incorporate the static self-energy into the FRG flow (rhs.~of Fig.\ref{fig4}a). Flowing into the symmetry broken phase~\cite{Profe2024} then yields the linear combination of order parameters minimizing the free energy.
The resulting charge imbalance of the nematic state is depicted in the inset of Fig.\ref{fig4}c.
Due to our choice for the interacting manifold, the reduction of $C_6$ symmetry down to $C_2$ is most prominently observed on the $(\gamma)$ Fermi sheet in Fig.\ref{fig4}c, while the other sheets only exhibit a proximity induced warping due to hybridisation with the $(d_{xy},d_{x^2-y^2})$ orbitals in good agreement with the ACMs of Fig.~\ref{fig2} and \ref{ExtDataFig3}.

To pin down the microscopic origin of this nematic charge order, we analyze the response of the system to the most prominent fluctuation channels on the kagome lattice by monitoring their instability scale throughout the FRG flow. In Fig.\ref{fig4}b, the fluctuation strength $\Theta$, extracted as the eigenvalue of the respective channel at a given wavevector $Q$, is depicted as a function of the energy cutoff $\Lambda$. We note that the $Q = 0$ charge order does already feature a sizable susceptibility compared to other orders involving translation symmetry breaking, i.e. with a finite $Q$ vector. This can be attributed to the fact that the long-range interaction supports charge imbalance between adjacent sites already at the mean-field level.
In the weak to intermediate coupling regime, this potential energy gain is usually dwarfed on the kagome lattice by the pronounced Van Hove scattering, that promotes a charge density with the nesting vector M~\cite{Wang2013, Kiesel2013} and a $2\times 2$ supercell reconstruction.
In most members of the 135 family, this charge modulation is indeed realized~\cite{Jiang2021, Neupert2022} due to the existence of low lying VHss of pure and mixed sublattice type in the electronic spectrum~\cite{Hu2022, Kang2022}.
Within the FRG formulation, this effect can be directly observed via a strong enhancement of the M-point charge susceptibility as the cutoff approaches the Fermi level~\cite{Wang2013, Zhan2024}.
In the case of \ce{CsTi3Bi5}, there is only a single $p$-type VHs close to the Fermi level and the chemical potential of the system is detuned from Van Hove filling by $\approx 0.22\,$eV, which substantially reduces the effect of Van Hove nesting.
With the accompanied sublattice interference still active~\cite{Kiesel2012}, the accessible phase space for on-site scattering in the Van Hove channel remains limited and the local repulsion can be dynamically overscreened by long-range interactions, which suppresses the formation of local moments and the emergence of magnetic orders.

In addition, the potential energy gain in the geometrically frustrated kagome geometry is also supplemented by a kinetic energy gain.
The charge imbalance pushes two VHss in the quasi-particle band structure up in energy, while the third one moves closer to the Fermi level, as shown in Fig.\ref{fig4}d. Thereby, the spectral weight is more efficiently moved away from the Fermi level than by the opening of a charge ordered gap at the VHs.
This combined effect of electronic kinematics and interactions leads to a spontaneous breaking of rotational symmetry by implying non-uniform sublattice occupation without enlarging the unit cell.

In this respect, \ce{CsTi3Bi5} provides a rare material realisation of a $d$-wave Pomeranchuk instability in its original sense, namely the spontaneous breaking of pointgroup symmetry by correlation effects resulting in a smooth deformation of the Fermi surface~\cite{Pomeranchuk1958, Chubukov2018}. This is directly reflected in the divergence of the associated scattering vertex in Fig.~\ref{fig4}b at the critical scale.
In the emergence of a PI, the non-trivial sublattice structure of the kagome lattice supplies additional degree of freedom and finite angular momentum channels, that allow to circumvent the constraints for PI in conventional Fermi liquid theory:
if there is only one electron species present in the system, a PI can only manifest as bond or current order~\cite{Nayak2000}, whose emergence is often hampered by charge and spin conservation rules~\cite{Kiselev2017, Wu2018}.
While these states accumulate a finite angular momentum in the relative coordinate between the constituents of the particle-hole pair, the obtained $d$-wave PI state in \ce{CsTi3Bi5} is characterised by a $l=2$ quantum number in the total angular momentum of the charge order parameter within the unit cell.
Therefore, a charge accumulation on one site and orbital can be compensated by a partial depletion on a different site, as apparent from Fig.\ref{fig4}c.

Notably, our results differ from the current opinion that the nematic charge order in \ce{CsTi3Bi5} is realised by a bond-type order with zero total momentum~\cite{Hu2023,Li2023}. That assessment was based on the observation of a momentum-dependent nature of the nematic Fermi surface distortion, which was incorrectly claimed to be unexplainable by a simple charge imbalance between the unit cell sites~\cite{Hu2023}. Diagrammatic calculations also suggested a bond-type order~\cite{Huang2024}. It is however important to highlight that this previous study includes only a subset of the functional renormalization group (FRG) diagrams, leading to a pronounced bias towards non-local charge orders and pronounced features from VHss.  Our unbiased FRG analysis reveals that these additional diagrams indeed favor the emergence of a site-type order in the $E_2$ irreducible representation and disfavor odd-parity bond order states ~\cite{Kiselev2017, Wu2018}. Consistency between the experimentally observed Fermi surface warping and a charge imbalance between different unit cell sites is achieved by the orbital structure of the PI, that induces a non-trivial momentum space structure of the nematic quasi-particle band structure via the orbital-to-band transformation.
The resulting Fermi surface reconstructions for both site- and bond-type order differ only sightly, underscoring the subtle complexities involved in characterizing intra-unit cell orders and the necessity of a synergetic approach via spectroscopic signatures like ARPES and detailed theoretical modelling.

In this study, we have elucidated the origin of the nematic character in CsTi$_3$Bi$_5$ by correlating high-resolution light-polarization angle-resolved photoemission spectroscopy data with ab initio-based many-body theoretical predictions. Our findings reveal that the nematicity is predominantly driven by electronic correlations through a complex orbital-selective mechanism and detuning of the chemical potential away from Van Hove filling. This work not only advances our understanding of nematic behavior within the 135 family of kagome materials but also provides critical insights into its underlying origins and manifestations. By establishing this connection, we paved the way for further exploration of electronic correlations and their implications in complex quantum materials.

On a broader perspective, our work suggests a generic propensity of PIs in systems with a non-trivial unit cell based on the frustrated hexagonal lattice, beyond \ce{CsTi3Bi5}. This seems to be witnessed by observations in related materials. 
In regard to other Ti-based kagome systems, a previous study on the sister compound \ce{RbTi3Bi5} employed ARPES measurements to investigate aspects of nematicity in kagome metals~\cite{Hu2023}. However, the observed changes were too subtle to draw definitive conclusions, with the results suggesting a potential contribution from $p_{x,y}$ and $d_{xy}$ orbitals to the nematicity, along with significant interband scattering, likely originating from the second hexagonal Fermi surface sheet and pockets at the M point. The lack of conclusive evidence in that study was likely due to the absence of measurements using multiple light polarizations. In contrast, our work leveraged light-polarization-dependent ARPES, which was crucial for revealing the orbital character of the band structure and extracting a clear signal from the ACMs, directly correlating with first-principles calculations.
Beyond the family of 135, in the kagome bilayer material \ce{ScV6Sn6}, a stripe-like nematic order was recently identified by STM measurements~\cite{Jiang2024}.
Due to the complexity of the electronic structure, the microscopic origin of this PI could not be identified, but was speculated to arise from Van Hove scattering effects.
The reduced complexity of \ce{CsTi3Bi5} allowed, for the first time, to pinpoint the physical origin of a nematic instability using rigorous many-body numerical techniques from ab initio calcultions.
In conclusions, our findings can serve as a guiding framework for understanding nematicity in other kagome metals, where multiple VHss near the Fermi level may cooperate to drive symmetry-breaking instabilities.

\bibliographystyle{prsty}
\bibliography{kagome.bib}

\begin{thebibliography}{10}

\bibitem{Kiesel2012}
M.~L. Kiesel and R. Thomale, Phys. Rev. B {\bf 86},  121105  (2012).

\bibitem{Hu2022}
Y. Hu {\it et~al.}, Nature Communications {\bf 13},  2220  (2022).

\bibitem{Kang2022}
M. Kang {\it et~al.}, Nature Physics {\bf 18},  301  (2022).

\bibitem{Kang2020}
M. Kang {\it et~al.}, Nature Communications {\bf 11},  4004  (2020).

\bibitem{Li2021}
M. Li {\it et~al.}, Nature Communications {\bf 12},  3129  (2021).

\bibitem{Liu2020}
Z. Liu {\it et~al.}, Nature Communications {\bf 11},  4002  (2020).

\bibitem{DiSante2023}
D. Di~Sante {\it et~al.}, Nature Physics {\bf 19},  1135  (2023).

\bibitem{Mielke2022}
C. Mielke {\it et~al.}, Nature {\bf 602},  245  (2022).

\bibitem{Yu2021}
L. {Yu} {\it et~al.}, arXiv e-prints  arXiv:2107.10714  (2021).

\bibitem{Guo2022}
C. Guo {\it et~al.}, Nature {\bf 611},  461  (2022).

\bibitem{Neupert2022}
T. Neupert, M.~M. Denner, J.-X. Yin, R. Thomale, and M.~Z. Hasan, Nature
  Physics {\bf 18},  137  (2022).

\bibitem{Chen2021}
H. Chen {\it et~al.}, Nature {\bf 599},  222  (2021).

\bibitem{Zhou2022}
S. Zhou and Z. Wang, Nature Communications {\bf 13},  7288  (2022).

\bibitem{Schwemmer2024}
T. Schwemmer {\it et~al.}, Phys. Rev. B {\bf 110},  024501  (2024).

\bibitem{Deng2024}
H. Deng {\it et~al.}, Nature {\bf 632},  775  (2024).

\bibitem{Jiang2021}
Y.-X. Jiang {\it et~al.}, Nature Materials {\bf 20},  1353  (2021).

\bibitem{Yin2022}
J.-X. Yin {\it et~al.}, Phys. Rev. Lett. {\bf 129},  166401  (2022).

\bibitem{Xu2022}
Y. Xu {\it et~al.}, Nature Physics {\bf 18},  1470  (2022).

\bibitem{Tuniz2023}
M. Tuniz {\it et~al.}, Communications Materials {\bf 4},  103  (2023).

\bibitem{Hu2023}
Y. Hu {\it et~al.}, Nature Physics {\bf 19},  1827  (2023).

\bibitem{Jiang2024}
Y.-X. Jiang {\it et~al.}, Nature Materials {\bf 23},  1214  (2024).

\bibitem{Drucker2024}
N. Drucker {\it et~al.}, arXiv preprint arXiv:2401.17141  (2024).

\bibitem{Nag2024}
P.~K. Nag {\it et~al.}, arXiv preprint arXiv:2410.01994  (2024).

\bibitem{Boehmer2022}
A.~E. B{\"o}hmer, J.-H. Chu, S. Lederer, and M. Yi, Nature Physics {\bf 18},
  1412  (2022).

\bibitem{kang2023}
M. Kang {\it et~al.}, Nature Materials {\bf 22},  186  (2023).

\bibitem{Le2024}
T. Le {\it et~al.}, Nature {\bf 630},  64  (2024).

\bibitem{Ortiz2019}
B.~R. Ortiz {\it et~al.}, Phys. Rev. Mater. {\bf 3},  094407  (2019).

\bibitem{Ortiz2020}
B.~R. Ortiz {\it et~al.}, Phys. Rev. Lett. {\bf 125},  247002  (2020).

\bibitem{Zhao_2021}
H. Zhao {\it et~al.}, Nature {\bf 599},  216  (2021).

\bibitem{Zheng2022}
L. Zheng {\it et~al.}, Nature {\bf 611},  682  (2022).

\bibitem{Li2023}
H. Li {\it et~al.}, Nature Physics {\bf 19},  1591  (2023).

\bibitem{Jiang2023}
Z. Jiang {\it et~al.}, Nature Communications {\bf 14},  4892  (2023).

\bibitem{Pomeranchuk1958}
I.~I. Pomeranchuk, Sov. Phys. JETP {\bf 8},  361  (1958).

\bibitem{Huang2024}
J. Huang, Y. Yamakawa, R. Tazai, T. Morimoto, and H. Kontani, arXiv preprint
  arXiv:2305.18093  (2023).

\bibitem{Saito2021}
Y. Saito {\it et~al.}, Nature {\bf 592},  220  (2021).

\bibitem{Rozen2021}
A. Rozen {\it et~al.}, Nature {\bf 592},  214  (2021).

\bibitem{Chichinadze2020}
D.~V. Chichinadze, L. Classen, and A.~V. Chubukov, Phys. Rev. B {\bf 102},
  125120  (2020).

\bibitem{Hossain2024}
M.~S. Hossain {\it et~al.}, arXiv preprint arXiv:2410.19636  (2024).

\bibitem{Yi2023}
X.-W. Yi, Z.-W. Liao, J.-Y. You, B. Gu, and G. Su, Research {\bf 6},  0238
  (2023).

\bibitem{Yang2023}
J. Yang {\it et~al.}, Nature Communications {\bf 14},  4089  (2023).

\bibitem{Wang2023}
Y. Wang {\it et~al.}, Chinese Physics Letters {\bf 40},  037102  (2023).

\bibitem{Damascelli2004}
A. Damascelli, Physica Scripta {\bf 2004},  61  (2004).

\bibitem{Damascelli2003}
A. Damascelli, Z. Hussain, and Z.-X. Shen, Rev. Mod. Phys. {\bf 75},  473
  (2003).

\bibitem{Fernandes2014}
R.~M. Fernandes, A.~V. Chubukov, and J. Schmalian, Nature Physics {\bf 10},  97
   (2014).

\bibitem{Nie2022}
L. Nie {\it et~al.}, Nature {\bf 604},  59  (2022).

\bibitem{Guo2024}
C. Guo {\it et~al.}, Nature Physics {\bf 20},  579  (2024).

\bibitem{Cichutek2022}
N. Cichutek, M. Hansen, and P. Kopietz, Phys. Rev. B {\bf 105},  205148
  (2022).

\bibitem{Salmhofer2001}
M. Salmhofer and C. Honerkamp, Progress of Theoretical Physics {\bf 105},  1
  (2001).

\bibitem{Metzner2012}
W. Metzner, M. Salmhofer, C. Honerkamp, V. Meden, and K. Sch\"onhammer, Rev.
  Mod. Phys. {\bf 84},  299  (2012).

\bibitem{Platt2013}
W.~H. C.~Platt and R. Thomale, Advances in Physics {\bf 62},  453  (2013).

\bibitem{Dupuis2021}
N. Dupuis {\it et~al.}, Physics Reports {\bf 910},  1  (2021), the
  nonperturbative functional renormalization group and its applications.

\bibitem{Beyer2022}
J. Beyer, J.~B. Profe, and L. Klebl, The European Physical Journal B {\bf 95},
  65  (2022).

\bibitem{Husemann2009}
C. Husemann and M. Salmhofer, Phys. Rev. B {\bf 79},  195125  (2009).

\bibitem{Lichtenstein2017}
J. Lichtenstein {\it et~al.}, Computer Physics Communications {\bf 213},  100
  (2017).

\bibitem{Profe2022}
J.~B. Profe and D.~M. Kennes, The European Physical Journal B {\bf 95},  60
  (2022).

\bibitem{Profe2024a}
J.~B. Profe, D.~M. Kennes, and L. Klebl, SciPost Phys. Codebases  26  (2024).

\bibitem{Kiesel2013}
M.~L. Kiesel, C. Platt, and R. Thomale, Phys. Rev. Lett. {\bf 110},  126405
  (2013).

\bibitem{Profe2024}
J.~B. Profe {\it et~al.}, arXiv preprint arXiv:2402.11916  (2024).

\bibitem{Wang2013}
W.-S. Wang, Z.-Z. Li, Y.-Y. Xiang, and Q.-H. Wang, Phys. Rev. B {\bf 87},
  115135  (2013).

\bibitem{Zhan2024}
J. Zhan {\it et~al.}, submitted.

\bibitem{Chubukov2018}
A.~V. Chubukov, A. Klein, and D.~L. Maslov, Journal of Experimental and
  Theoretical Physics {\bf 127},  826  (2018).

\bibitem{Nayak2000}
C. Nayak, Phys. Rev. B {\bf 62},  4880  (2000).

\bibitem{Kiselev2017}
E.~I. Kiselev, M.~S. Scheurer, P. W\"olfle, and J. Schmalian, Phys. Rev. B {\bf
  95},  125122  (2017).

\bibitem{Wu2018}
Y.-M. Wu, A. Klein, and A.~V. Chubukov, Phys. Rev. B {\bf 97},  165101  (2018).

\bibitem{growth}
Zeitschrift für Naturforschung B {\bf 77},  i  (2022).

\bibitem{PhysRevB.59.1743}
K. Koepernik and H. Eschrig, Phys. Rev. B {\bf 59},  1743  (1999).

\bibitem{PhysRevB.45.13244}
J.~P. Perdew and Y. Wang, Phys. Rev. B {\bf 45},  13244  (1992).

\bibitem{PhysRev.52.191}
G.~H. Wannier, Phys. Rev. {\bf 52},  191  (1937).

\bibitem{RevModPhys.34.645}
G.~H. Wannier, Rev. Mod. Phys. {\bf 34},  645  (1962).

\bibitem{PhysRevB.59.1758}
G. Kresse and D. Joubert, Phys. Rev. B {\bf 59},  1758  (1999).

\bibitem{PhysRevB.54.11169}
G. Kresse and J. Furthm\"uller, Phys. Rev. B {\bf 54},  11169  (1996).

\bibitem{PhysRevB.50.17953}
P.~E. Bl\"ochl, Phys. Rev. B {\bf 50},  17953  (1994).

\bibitem{PhysRevB.46.6671}
J.~P. Perdew {\it et~al.}, Phys. Rev. B {\bf 46},  6671  (1992).

\bibitem{PhysRevA.38.3098}
A.~D. Becke, Phys. Rev. A {\bf 38},  3098  (1988).

\bibitem{PhysRevB.28.1809}
D.~C. Langreth and M.~J. Mehl, Phys. Rev. B {\bf 28},  1809  (1983).

\bibitem{PhysRevLett.77.3865}
J.~P. Perdew, K. Burke, and M. Ernzerhof, Phys. Rev. Lett. {\bf 77},  3865
  (1996).

\bibitem{Wu2021}
X. Wu {\it et~al.}, Phys. Rev. Lett. {\bf 127},  177001  (2021).

\bibitem{DiSante2023a}
D. Di~Sante {\it et~al.}, Phys. Rev. Res. {\bf 5},  L012008  (2023).

\end{thebibliography}

\section{Methods}

\subsection{Bulk single-crystal synthesis} Single crystals of \ce{CsTi3Bi5} were grown using a conventional flux-based growth technique, as described previously \cite{growth}. The elemental ingredients for synthesis included Cs (liquid, Alfa 99.98$\,\%$), Ti (powder, Alfa 99.9$\,\%$), and Bi (shot, Alfa 99.999$\,\%$). These materials were loaded into a tungsten carbide milling vial in a stoichiometric ratio of 1:1:6 and milled for an hour under an argon atmosphere to obtain precursor powder. After milling, the precursor powder was transferred into an alumina crucible and sealed in a separate stainless-steel tube. The samples were heated at 900$^\circ\,$C for 10 h and then cooled at 3$^\circ\,$C/hr to 500$^\circ\,$C. Once the growth period was over, the tube was broken inside a glove box filled with Ar-gas. The shiny plate-like single crystals were separated gently, and stored inside a box filled with an inert gas atmosphere.
As also shown in previous studies~\cite{Li2023}, the single domains are extremely large in CsTi$_3$Bi$_5$ compared to those of the vanadium-based sister compounds. This aspect permits the extraction and analysis of autocorrelation maps from the ARPES spectra of Fermi surface.

\subsection{ARPES experiments} The samples were cleaved in ultrahigh vacuum (UHV) at the pressure of $1 \times 10^{-10}\,$mbar. The ARPES data were acquired at the CASSIOPEE end station of the synchrotron radiation source SOLEIL (Paris, France). The energy and momentum resolutions were better than 10 meV and 0.018 \AA$^{-1}$, respectively. The temperature of the measurements was kept constant throughout the data acquisitions (15 K). The Fermi surfaces were collected by rotating the angle orthogonal to the analyser slit, keeping the samples in the centre of rotation. To exclude possible matrix elements effects, several light polarizations, geometries, and photon energies were used for the data acquisition, yielding always consisting results.

\subsection{First-principles calculations}
Bulk electronic structure calculations were performed using the full-potential local-orbital (FPLO) code (v.21.00-61) \cite{PhysRevB.59.1743}. The unit cell has lattice constants of 5.82709 Å, 5.82709 Å, and 9.93612 Å. The exchange-correlation energy was parametrized within the local density approximation, following the Perdew-Wang 92 formulation \cite{PhysRevB.45.13244}. A $12\times12\times12$ $k$-grid was used to sample the Brillouin Zone, and the tetrahedron method was employed for integration. Calculations were performed in both the full-relativistic and non-relativistic frameworks.\\
A subsequent 36 bands Wannier functions model \cite{PhysRev.52.191, RevModPhys.34.645}, with symmetries accurately implemented, was constructed considering projections onto the following states: Caesium $6s$; Titanium $3d_{z^2}$, $3d_{xz}$, $3d_{yz}$, $3d_{x^2 - y^2}$, $3d_{xy}$; and Bismuth $6p_z$, $6p_x$, $6p_y$, $6s$.\\
Surface electronic structure calculations were performed  using the Vienna Ab initio Simulation Package (VASP) \cite{PhysRevB.59.1758, PhysRevB.54.11169}, using the projector augmented wave (PAW) method \cite{PhysRevB.50.17953}. Exchange and correlation effects have been handled using the generalized gradient approximation (GGA) \cite{PhysRevB.46.6671,PhysRevA.38.3098,PhysRevB.28.1809} within the Perdew-Burke-Ernzerhof (PBE) approach \cite{PhysRevLett.77.3865}. The truncation of the basis set was set by a plane-wave cutoff of 500 eV and a $12\times12\times1$ $k$-grid was used.  

\subsection{Many-Body calculations}

To analyse the symmetry breaking propensities of \ce{CsTi3Bi5}, we employed the functional renormalization group (FRG) in the truncated unity formalism implemented in the divERGe library~\cite{Profe2024a}.
Despite effects of the out-of-plane dispersion on our FRG results are very weak (cf. Supplementary Information II), we evaluate the scattering vertex in the respective channels on a $24 \times 24 \times 6$ mesh in the full 3D Brillouin zone, with an additional refinement of $12 \times 12 \times 5$ for the integration of the internal loops in the flow equations. We include real space form-factors up to fourth nearest neighbour, that have shown to produce converged results on the kagome Hubbard model~\cite{Profe2024}. 
\\
The FRG interpolates between the full interacting model and an effective low energy theory close the Fermi level by adding a regulator $\Lambda$ to the bare Green's function, that recovers the non-interaction limit as $\Lambda \rightarrow \infty$, and separates the electronic states into fast and slow modes~\cite{Metzner2012, Platt2013}.
As the cutoff $\Lambda$ is successively lowered, the fast modes are integrated out and provide the screening background for the effective interaction of the remaining degrees of freedom via the mutual electron-electron interactions between fast and slow modes.
The change of the effective $n-$particle interaction as a function of the cutoff is best expressed in an infinite hierarchy of coupled differential flow equations for the $2n-$particle vertex, that is conventionally truncated within one loop order, i.e. at second order in the interactions, and gives way to the FRG flow equations depicted in Fig.\ref{fig4}a.
By monitoring the evolution of the vertex throughout the flow, the FRG provides a transparent way to pinpoint the origin of a symmetry breaking transition. Its implicit diagrammatic resummation is unbiased in the sense that it includes all ladder diagrams of the particle-particle (P), direct particle-hole (D), and crossed particle-hole (C) channel, as well the leading order vertex corrections between them. This cross talk between the different diagrammatic channels proves quintessential to approach cooperative and competing instability phenomena characteristic for the highly frustrated kagome lattice~\cite{Schwemmer2024, Profe2024}.
\\
We mimic the mutual electron-electron interaction within the Ti $d$-orbital manifold at the bare level by the two particle vertex
\begin{equation}
\begin{split}
    \hat{H}_I = &
       \, U \sum_{i o}{\hat{n}_{i o \uparrow}\hat{n}_{i o \downarrow}}
     + \, U^\prime \sum_{i o_1 < o_2 \sigma \sigma^\prime}{\hat{n}_{i o_1 \sigma}\hat{n}_{i o_2 \sigma^\prime}} \\
     + &~ \, J \sum_{i o_1 < o_2 \sigma \sigma^\prime}
        {\hat{c}^\cre_{i o_1 \sigma} \hat{c}^\cre_{i o_2 \sigma^\prime}
        \hat{c}^\ann_{i o_1 \sigma^\prime} \hat{c}^\ann_{i o_2 \sigma}}
     + \, J^\prime \sum_{i o_1 \neq o_2 }
        {\hat{c}^\cre_{i o_1 \uparrow} \hat{c}^\cre_{i o_1 \downarrow}
        \hat{c}^\ann_{i o_2 \downarrow} \hat{c}^\ann_{i o_2 \uparrow}} \\
     + &~ V_1 \sum_{\langle i, j \rangle, o_1 o_2 \sigma\sigma'}{\hat{n}_{j o_1 \sigma}\hat{n}_{i o_2 \sigma'}} 
     + V_2 \sum_{\langle \langle i, j \rangle \rangle, o_1 o_2 \sigma\sigma'}{\hat{n}_{j o_1 \sigma}\hat{n}_{i o_2 \sigma'}}\, ,
\end{split}
\label{eqn:interaction}
\end{equation}
where $\hat{n}_{i o \sigma} = \hat{c}^{\dagger}_{i o \sigma}\hat{c}_{i o \sigma}$  is the fermionic number operator on site $i$ with orbital $o$ and spin $\sigma$.
$\langle i, j \rangle$ and $\langle \langle i, j \rangle \rangle$ indicate a summation over nearest neighbor (NN) and second nearest neighbor (NNN) sites respectively.
\\
In accordance with the dominant role of the in-plane \ce{Ti} $d$-orbitals in the low-energy kinematics described above (Fig.\ref{fig3}d,e), we choose the interacting manifold as the in-plane \ce{Ti} $d$-orbitals, while all other orbitals only provide the screening background for the two particle interaction in this channel.
We have confirmed the validity of our results via calculations with an interacting manifold containing the full $d$-shell (cf. Supplementary Information III).
\\
We use the universal Kanamori relations for the onsite interaction tensor $U = U^\prime + 2 J$ and $J = J^\prime$~\cite{Wu2021} with $ J = 0.8\,$eV and set $U = 4\,$eV, $V_1 = 1.5\,$eV, and $V_2 = 0.5\,$eV for all data provided in the main paper by employing the universal decay behaviour for density-density interactions in the 135 family obtained from constrained RPA calculations~\cite{DiSante2023a}.
All our results are robust against moderate changes of the interaction profile.

\section{Data availability}

All data needed to evaluate the conclusions in the paper are present in the Article and its Supplementary Information. Additional data are available from the corresponding author upon reasonable request.

\section{Acknowlegments}
We acknowledge SOLEIL for provision of synchrotron radiation facilities under proposal No. 20231813. M.D., R.T. and G.S. are supported by the Deutsche Forschungsgemeinschaft (DFG, German Research Foundation) through Project-ID 258499086 - SFB 1170 and through the W\"urzburg-Dresden Cluster of Excellence on Complexity and Topology in Quantum Matter - ct.qmat Project-ID 390858490 - EXC 2147. F.M. is grateful for the project funded by the European Union – NextGenerationEU, M4C2, within the PNRR project NFFA-DI, CUP B53C22004310006, IR0000015. A.C. acknowledges support from PNRR MUR project PE0000023-NQSTI. A.C., R.T., G.S. and D.D.S. acknowledge the Gauss Centre for Supercomputing e.V. (https://www.gauss-centre.eu) for funding this project by providing computing time on the GCS Supercomputer SuperMUC-NG at Leibniz Supercomputing Centre (https://www.lrz.de). M.D., L.K. and R.T. are grateful for HPC resources provided by the Erlangen National High Performance Computing Center (NHR@FAU) of the Friedrich-Alexander-Universit\"at Erlangen-N\"urnberg (FAU), that were used for the FRG calculations. NHR funding is provided by federal and Bavarian state authorities. NHR@FAU hardware is partially funded by the DFG - 440719683. I.Z. gratefully acknowledges the support from the National Science Foundation (NSF), Division of Materials Research 2216080.
S.D.W., B.R.O., and G.P. gratefully acknowledge support via the UC Santa Barbara NSF Quantum Foundry funded via the Q-AMASE-i program under award DMR-1906325. We thank C.A. Baum and H. Hohmann for valuable feedback on the FRG calculations.

\section{Author contributions}

C.B., F.M., F. B., P. L. F., A. D. V. conducted the ARPES experiments and C.B. and F.M. analysed the experimental data. M.D., L.K. and A.C. carried out the theoretical analysis in consultation with D.D.S., G.S. and R.T. S.D.W., B.O. and G.P. performed the crystal synthesis and structural characterization. F.M., C.B., M.D. and D.D.S. wrote the paper with contributions from all authors. All authors discussed the results, interpretation and conclusion.

\section{Competing interests}
The authors declare no competing interests.

\newpage

\counterwithin*{figure}{part}

\stepcounter{part}

\renewcommand{\thefigure}{Extended Data Fig.\arabic{figure}}

\begin{figure*}[t]
	\centering
	\includegraphics[width=\textwidth]{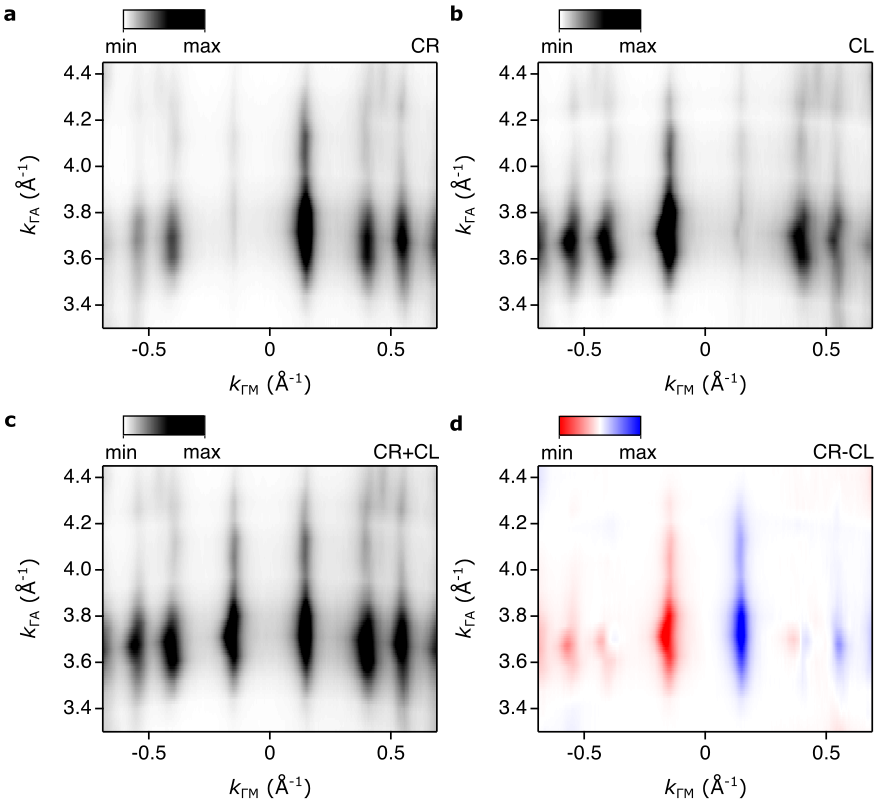}
    \caption{\textbf{$k_z$ versus $k_{||}$ at the Fermi level.} . The $k_z$-maps are obtained by scanning h$\nu$ from 40 eV to 70 eV with both \textbf{a} Circular Right (CR) and \textbf{b} Circular Left (CL) helicities. \textbf{c} unpolarised light obtained by summing up the $k_z$ maps collected with right- and left-handed circularly polarised light and \textbf{d} the respective circular dichroism. Sample is oriented with the $\Gamma$-M direction along the analyser's slit.}
    \label{ExtDataFig1}
\end{figure*}

\newpage

\begin{figure*}[t]
	\centering
	\includegraphics[width=\textwidth]{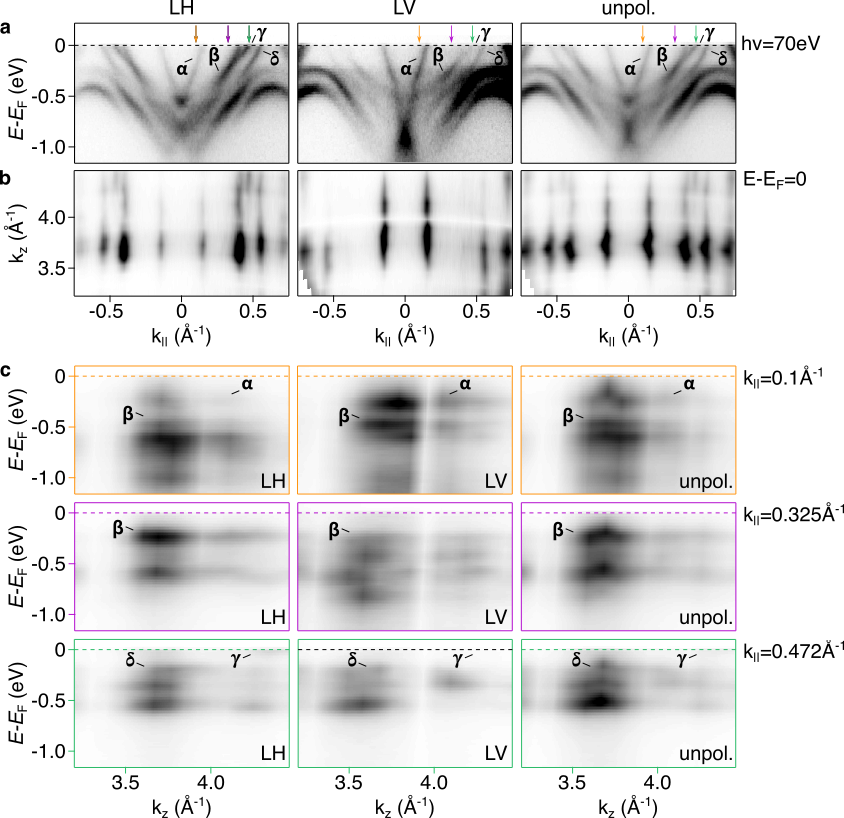}
    \caption{\textbf{Band structure two-dimensional character.} \textbf{a}, Band dispersion probed with $70$ eV photon energy. \textbf{b}, $k_z$-maps at the Fermi level obtained by scanning $h\nu$ from $32.5$ eV to $70$ eV. \textbf{c}, The absence of striking band dispersion along $k_z$ has been sampled at different $k_{||}$ points (marked by coloured arrows in panel \textbf{a}) is consistent with the calculated two-dimensional character of the Fermi contour. To rule out geometrical and matrix element effects, we used both linear horizontal, linear vertical and unpolarised (CR+CL) light polarisations. Sample is oriented with the $\Gamma$-M direction along the analyser's slit. The relevant bands building the Fermi contours are labelled following the same notation of the main text.}
	\label{ExtDataFig1bis}
\end{figure*}

\newpage

\begin{figure*}[t]
	\centering
	\includegraphics[width=\textwidth]{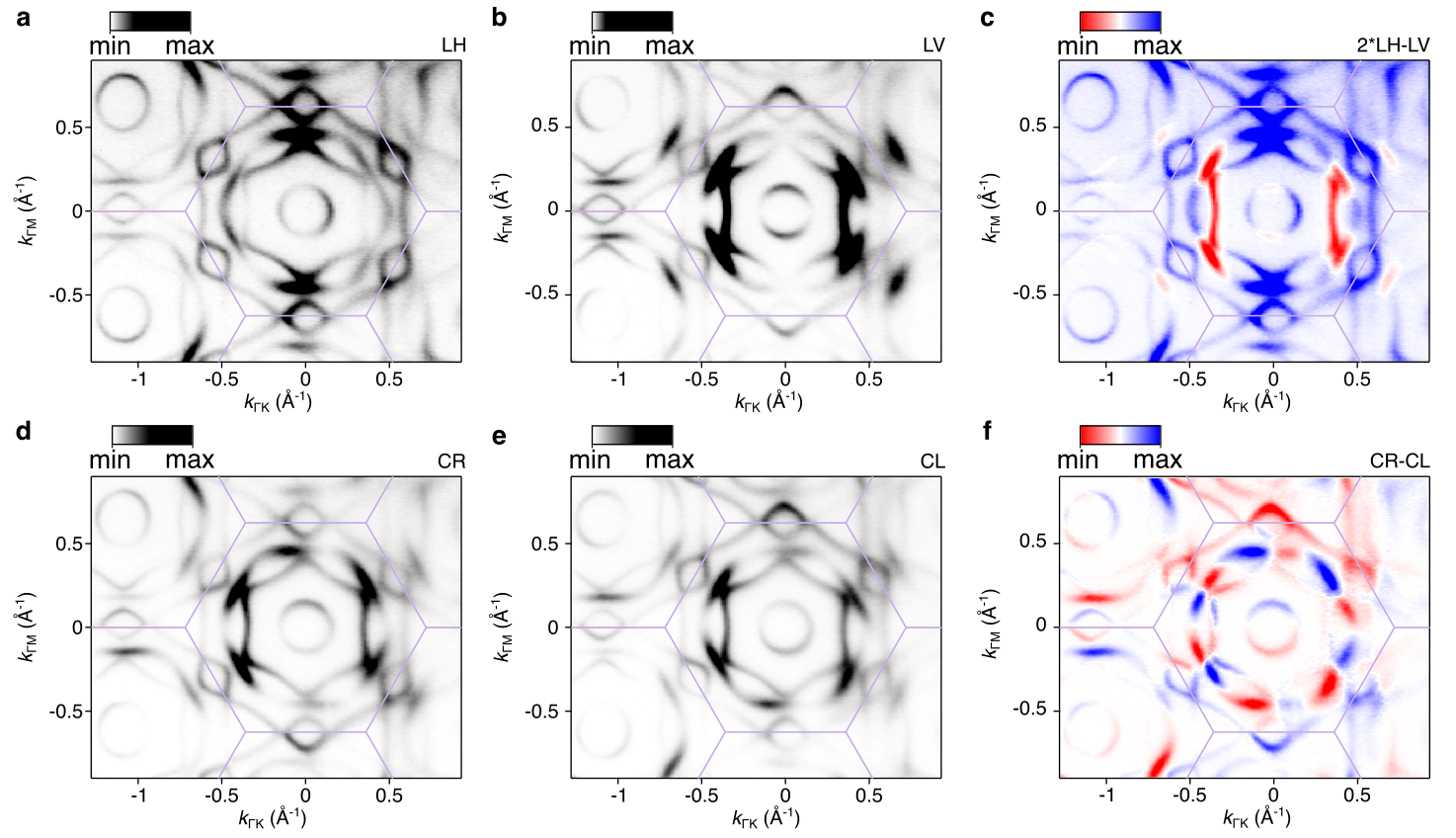}
	\caption{\textbf{Polarisation dependence of the Fermi contours.} Fermi surface contours measured in the $\Gamma$KM plane (h$\nu$=65 eV) with different polarisations: \textbf{a} linear horizontal, \textbf{b} linear vertical and \textbf{c} the resulting linear dichroism. \textbf{d} circular right, \textbf{e} circular left and \textbf{f} the resulting circular dichroism.}
	\label{ExtDataFig2}
\end{figure*}

\newpage

\begin{figure*}[t]
	\centering
	\includegraphics[width=0.9\textwidth]{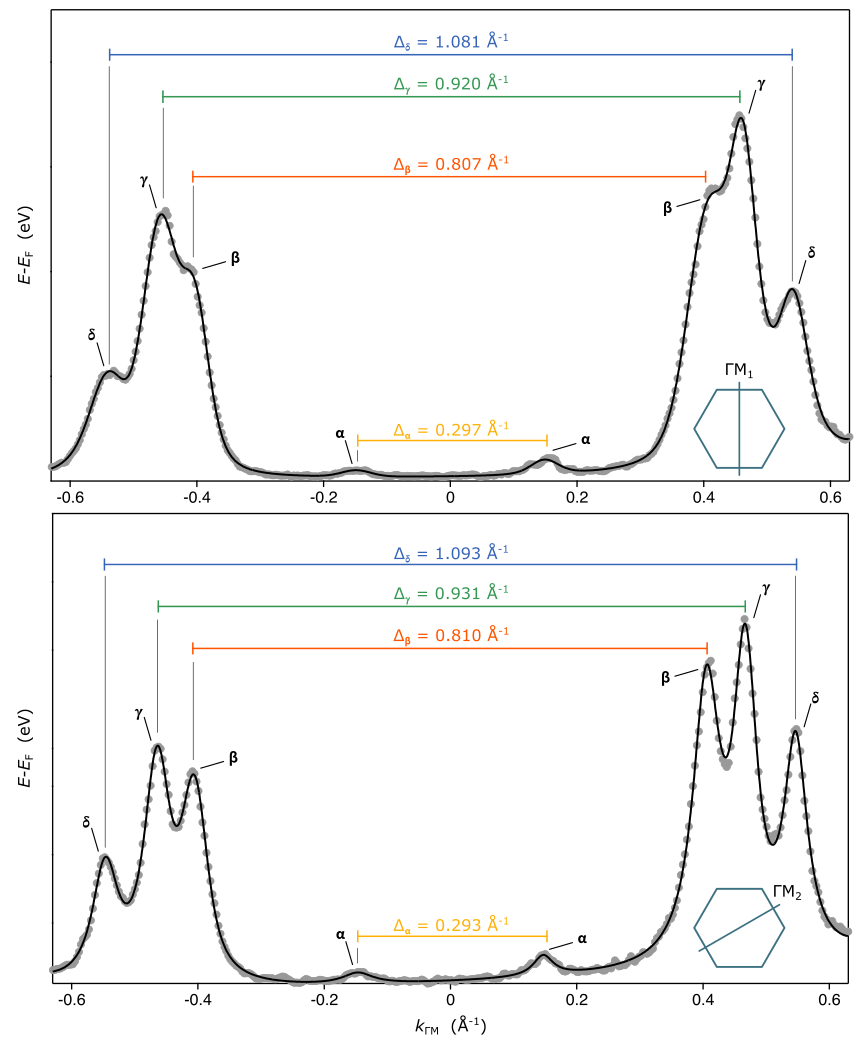}
    \caption{\textbf{Lorentzian fit at the Fermi level.} Momentum distribution curves (MDCs) extracted at the Fermi level along two inequivalent $\Gamma$-M directions (LH) to show the changes in the fermi momenta of ($\gamma$) and ($\delta$) bands. Grey dots are the data points, and the black solid line is the respective fit. The fit, to capture all the bands, was done on data collected by using LH polarization.}
    \label{ExtDataFig3}
\end{figure*}

\newpage

\begin{figure*}[t]
	\centering
	\includegraphics[width=\textwidth]{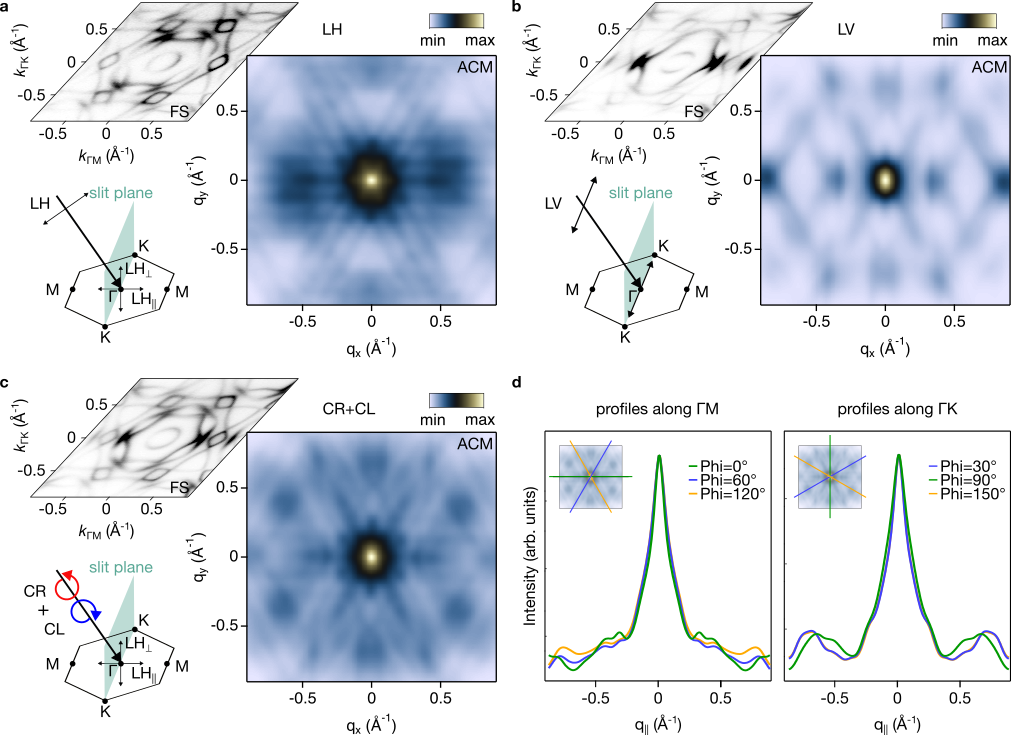}
	\caption{\textbf{Electronic nematicity of the Fermi contour.} Analogous of Fig.\ref{fig2} but with $\Gamma$-K high-symmetry direction aligned along the analyser's slit. \textbf{a}, Autocorrelation map (ACM) extracted at the Brillouin zone centre with h$\nu$ = 65 eV for linearly horizontal polarised light, \textbf{b} linear vertical polarisation and \textbf{c} unpolarised light obtained by summing up spectra collected with right- and left-handed circularly polarised light. The cartoons illustrate the experimental geometry and vector projections. \textbf{d}, Azimuthal profiles (extracted from the ACM in \textbf{c}) emphasize the reduced C$_2$ symmetry.}
	\label{ExtDataFig4}
\end{figure*}

\newpage

\begin{figure*}[t]
	\centering
	\includegraphics[width=\textwidth]{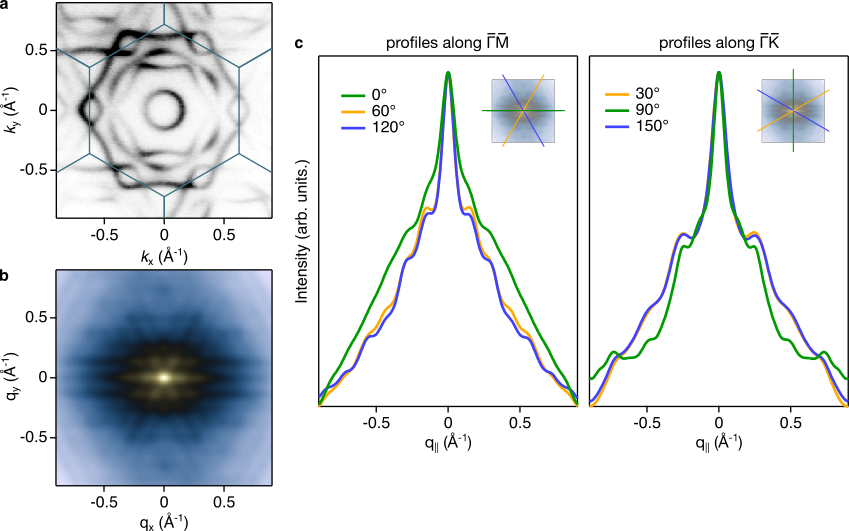}
     \caption{\textbf{Electronic nematicity at finite $k_z$.} \textbf{a}, Fermi surface contour obtained with $75$ eV linear horizontal polarisation light, \textit{i.e.} away from the Brillouin zone center $k_z=0$. The sample is oriented with the $\Gamma$-K direction along the analyser's slit. \textbf{b}, The associated autocorrelation map. \textbf{c}, Azimuthal profiles extracted from the ACM in \textbf{c}, supporting the reduced nematic C$_2$ symmetry.}
	\label{ExtDataFig6}
\end{figure*}

\end{document}